\documentclass[twocolumn,prd,superscriptaddress,showpacs,floatfix,%
preprintnumbers,nofootinbib,amsfont]{revtex4}

\usepackage{epsfig}

\usepackage[dvips]{color}
\definecolor{Black}{named}{Black}
\definecolor{Blue}{named}{Blue}
\definecolor{Red}{named}{Red}

\begin{document}

\title{Exploiting the directional sensitivity of the
Double Chooz near detector}

\author{Kathrin A.~Hochmuth}\email{hochmuth@ph.tum.de}
\affiliation{Max-Planck-Institut f\"ur Physik
(Werner-Heisenberg-Institut), F\"ohringer Ring 6, 80805 M\"unchen,
Germany}

\author{Manfred Lindner}\email{lindner@mpi-hd.mpg.de}
\affiliation{Max-Planck-Institut f\"ur Kernphysik,
Saupfercheckweg~1, 69117 Heidelberg, Germany}

\author{Georg G.~Raffelt}\email{raffelt@mppmu.mpg.de}
\affiliation{Max-Planck-Institut f\"ur Physik
(Werner-Heisenberg-Institut), F\"ohringer Ring 6, 80805 M\"unchen,
Germany}

\date{13 April 2007}

\preprint{MPP-2007-24}

\begin{abstract}
In scintillator detectors, the forward displacement of the neutron
in the reaction $\bar\nu_e+p\to e^++n$ provides neutrino directional
information as demonstrated by the CHOOZ reactor experiment with
2,500 events. The near detector of the forthcoming Double Chooz
experiment will collect $1.6\times10^5$ events per year, enough to
determine the average neutrino direction with a $1\,\sigma$
half-cone aperture of $2.3^\circ$ in one year. It is more difficult
to separate the two Chooz reactors that are viewed at a separation
angle $\phi=30^\circ$. If their strengths are known and
approximately equal, the azimuthal location of each reactor is
obtained with $\pm6^\circ$ ($1\,\sigma$) and the probability of
confusing them with a single source is less than 11\%. Five
year's data reduce this ``confusion probability'' to less than
0.3\%, i.e., a $3\,\sigma$ separation is possible. All of these
numbers improve rapidly with increasing angular separation of the
sources. For a setup with $\phi=90^\circ$ and one year's data, the
azimuthal $1\,\sigma$ uncertainty for each source decreases to
$\pm3.2^\circ$. Of course, for Double Chooz the two reactor
locations are known, allowing one instead to measure their
individual one-year integrated power output to $\pm11\%$
($1\,\sigma$), and their five-year integrated output to
$\pm4.8\%$~($1\,\sigma$).
\end{abstract}


 \maketitle

\section{Introduction}                        \label{sec:introduction}

The search for the neutrino mixing angle $\theta_{13}$ has led to a
number of proposals for reactor neutrino experiments, where
anti-neutrinos are registered in liquid scintillator detectors by the
inverse $\beta$ decay
\begin{equation}\label{det}
\bar{\nu}_e+p\rightarrow e^+ + n\,.
\end{equation}
In this reaction the neutron is scattered preferentially in the
forward direction so that it retains some memory of the neutrino's
initial direction, an effect first observed in experiments at the
G\"osgen reactor complex~\cite{goesgen}. By reconstructing the
vertices of the positron and neutron absorptions, one obtains an
image of the neutrino source.

The CHOOZ experiment demonstrated the feasibility of this approach
in that 2,500 events were enough to locate the source within a
$1\,\sigma$ half-cone aperture of $18^\circ$~\cite{chooz,chooz19}.
In principle, this method also allows one to determine the location
of a galactic supernova explosion~\cite{chooz,chooz19} and the
distribution of anti-neutrinos emitted by the natural radioactive
elements in the Earth~\cite{lena1,lena2}, although in practice these
applications are severely limited by the relatively small number of
events.

Here we investigate the potential of future reactor experiments to
exploit the same effect, but with much larger statistics. In
particular, the upcoming Double Chooz experiment~\cite{proposal} will be a first important test of the principles of directional measurements, which will explore the requirements for future large volume detectors. Double Chooz will operate two nearly identical detectors at distances of roughly
280~m and 1,050~m, respectively. The near detector will register
$1.6\times10^5$ events per year, vastly exceeding
the exposure of the CHOOZ experiment that was located at the far
site of Double Chooz and had only a short data-taking period. Thus one
year of Double Chooz data correspond to 64 times the CHOOZ exposure and hence to an
8-fold improved angular resolution, implying that the neutrino
source can be located within a $1\,\sigma$ half-cone aperture
of~$2.3^\circ$.

The Chooz nuclear power plant consists of two reactors that are
viewed by the near detector at an angular separation $\phi=30^\circ$
so that one may well wonder if it is possible to separate the
``neutrino images'' of the two sources and/or to monitor their
individual neutrino and thus power output. In spite of the
impressive single-source directional sensitivity, this is not
entirely obvious or trivial. Even though the average neutrino
direction can be determined very well, separating two very blurred
neutrino images is not simple even with the large event rate at the
Double Chooz near detector.

In Sec.~\ref{sec:setup} we outline the Double Chooz experiment and specify our simplifying assumptions where we use the
previous CHOOZ detector properties as our benchmark. In
Sec.~\ref{sec:analytical} we provide simple analytic estimates
before turning in Sec.~\ref{sec:maxlike} to a detailed Monte Carlo
analysis of the Double Chooz setup and a hypothetical setup with a larger separation
angle that could be of relevance to future experiments. We conclude
in Sec.~\ref{sec:conclusions}.

\section{Experimental Setup}
\label{sec:setup}

Both Double Chooz detectors will consist of about 10~tons of
Gadolinium-loaded scintillator, with a Gd concentration of 0.1\%.  The
final-state neutron of reaction Eq.~(\ref{det}) is captured by a Gd
nucleus with an efficiency of 90\%, releasing 2--3 $\gamma$ rays with
a total energy of about 8~MeV. The directional sensitivity of such
detectors rely on the forward displacement of the final-state neutron
in Eq.~(\ref{det}) relative to the location of the final-state
positron annihilation. Taking into account scattering and
thermalization, an average displacement $\ell=1.7~{\rm cm}$ with an
rms uncertainty of approximately 2.4~cm for the $x$-, $y$- and
$z$-directions was calculated~\cite{vogel}. Experimentally, the CHOOZ
experiment found $\ell=1.9 \pm 0.4$~cm~\cite{chooz}. To be specific we
will use
\begin{equation}
\ell=1.9~{\rm cm}\,,
\end{equation}
representing an average over the incoming neutrino energies and the
outgoing neutron directions. The large spatial distribution of the
neutron absorption points as well as the even larger uncertainty of
the neutron and positron event reconstruction imply that indeed only
the average of the neutron displacement matters here.

The experimental output used for the directional information is a
set of reconstructed displacement vectors ${\bf r}_i$ between the
positron-annihilation and neutron-capture events, where
$i=1,\ldots,N$. In the Double Chooz near detector we have in one
year
\begin{equation}
N=1.6\times10^5
\end{equation}
events, originating from both reactor cores together. Unless
otherwise stated we will always use this event number in our
numerical estimates.

Our main simplifying assumption is that for a single neutrino source
the distribution of the displacement vectors is a Gaussian with equal
width $L$ in each direction.  In the first CHOOZ paper addressing the
neutrino imaging of a reactor~\cite{chooz}, an rms uncertainty for the
neutron event reconstruction of 17--17.5~cm was given in their
Fig.~2. Their Fig.~3 implies $L=19$--20~cm and thus a positron event
reconstruction uncertainty of 8--10~cm, in agreement with a similar
result of the Borexino collaboration~\cite{borx}. In a later CHOOZ
publication~\cite{chooz19}, the rms uncertainty for the neutron event
reconstruction was given as 19~cm. Assuming 9~cm for the average
positron reconstruction uncertainty leads to
\begin{equation}
L=21~{\rm cm}
\end{equation}
that we will use as our benchmark value. With the planned
photomultiplier coverage in Double Chooz one does not expect to
improve on this value~\cite{privcom} so that the CHOOZ
characteristics provide a realistic estimate.

Given a Gaussian distribution of width $L$, its center of gravity
can be determined with a $1\,\sigma$ precision $L/\sqrt{N}$.
Therefore, the $1\,\sigma$ uncertainty for the angular location of a
single source is $(L/\ell)/\sqrt{N}$ that applies separately to the
azimuthal and zenith angle. One year's data provide an angular
uncertainty of $\pm1.58^\circ$. If both the azimuthal and zenith
angle are not known, the corresponding $1\,\sigma$ half-cone
aperture for the source location is $2.4^\circ$. Scaling this to
2,500 events leads to $19.2^\circ$, corresponding reasonably well to
the CHOOZ value of $18^\circ$~\cite{chooz}.

\begin{figure}[ht]
\includegraphics[width=0.6\columnwidth]{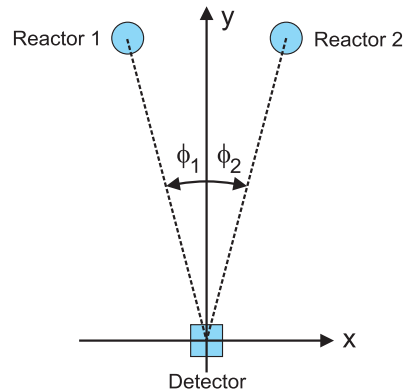}
\caption{Geometric setup.}\label{fig:setup}
\end{figure}

We will investigate a situation with two reactors that, together
with the detector, define the $x$-$y$-plane of our coordinate
system. The $y$-axis is taken to point from the detector towards the
reactors (Fig.~\ref{fig:setup}). The detector views the reactors
with azimuthal angles $\phi_1$ and $\phi_2$ relative to the
$y$-direction. We have chosen $\phi_1$ to have a positive and
$\phi_2$ a negative sense of rotation. The common zenith angle
$\theta$ of both reactors is measured against the $x$-$y$-plane. We
use $\theta=0^\circ$ for the true reactor locations, but in general
$\theta$ can be a fit parameter (Sec.~\ref{sec:tilt}).

With this geometric setup, the two reactor sources produce a
normalized distribution of positron-neutron displacement vectors
${\bf r}=(x,y,z)$ of
\begin{widetext}
\begin{equation}\label{eq:distribution}
 f({\bf r})=\frac{1}{(2\pi)^{3/2}\,L^3}\,
 \exp\left[-\frac{(z+\ell\sin\theta)^2}{2L^2}\right]\,
 \sum_{i=1}^2
 b_i\exp\left[
 -\frac{(x+\ell\cos\theta\sin\phi_i)^2+
 (y+\ell\cos\theta\cos\phi_i)^2}{2L^2}\right]\,.
\end{equation}
\end{widetext}
Here, $b_i$ with $b_1+b_2=1$ represent the individual reactor
contributions to the total event number $N$. Moreover, we define the
separation angle $\phi=|\phi_1-\phi_2|$ and the average of the neutrino direction
\begin{equation}
\phi_{c}=b_1\phi_1+b_2\phi_2.
\end{equation}

\section{Analytic Estimates}
\label{sec:analytical}

\subsection{Width {\boldmath $L$} of displacement-vector
distribution}
\label{sec:width}

The width $L$ of the distribution of the reconstructed
positron-neutron displacement vectors ${\bf r}$ can be determined from
the Double Chooz experiment itself. Assuming that the detector
response is spherically symmetric, one can extract $L$ from the
$z$-distribution $f_z(z)$ of the displacement vectors. This
distribution is equivalent to that from a single source. The
distribution of $L$ for many realizations, each with $N\gg 1$, is
essentially Gaussian~with
\begin{equation}
\frac{\sigma_L}{L}=\frac{1}{\sqrt{2N}}\,.
\end{equation}
One year's data provide $L$ with a fractional precision of
$1.7\times10^{-3}$ so that its uncertainty is negligible for our
further discussion. In addition, one could test deviations from the
assumed Gaussianity of the distribution.

\subsection{Average neutron displacement {\boldmath $\ell$}}
\label{sec:displacement}

The average neutron forward displacement $\ell$ can be extracted
from the $y$-distribution of the displacement vectors. Assuming a
symmetric setup with $\phi_1=-\phi_2=\beta$, $f({\bf r})$ factorizes
as $f_x(x) f_y(y) f_z(z)$ and $f_y(y)$ is independent of the
relative reactor strengths. The average and variance are $\bar
y=\ell\cos\beta$ and $\langle y^2-\bar y^2\rangle=L^2$ so that
\begin{eqnarray}
 \ell&=&\frac{\bar y}{\cos\beta}\,,
 \nonumber\\
 \frac{\sigma_\ell}{\ell}&=&\frac{L}{\ell}\,
 \frac{1}{\cos\beta\,\sqrt{N}}\,.
\end{eqnarray}
In Double Chooz we have $\phi=2\beta=30^\circ$ or $\cos\beta=0.966$.
For the purpose of determining $\ell$, the two reactors almost act
as a single source even at the near detector. After one year the
$1\,\sigma$ uncertainty will be $\pm2.9\%$. Scaled to 2,500 events,
this forecast corresponds reasonably well to the $\ell$ uncertainty
of $\pm20\%$ found by CHOOZ.

\subsection{Relative reactor strength}

As a first nontrivial application we address the question of how
well one can monitor the relative reactor strengths. With
$\phi_1=-\phi_2=\beta$, only the $x$-distribution carries
information on $b_1$ and $b_2$, and in particular
\begin{eqnarray}
 \bar x&=&(b_1-b_2)\,\ell\sin\beta\,,
 \nonumber\\
 \langle x^2-\bar x^2\rangle&=&
 L^2+[(1-(b_1-b_2)^2]\,(\ell\sin\beta)^2\,.
\end{eqnarray}
The variance is very close to $L^2$ because $L\gg\ell$ so that
\begin{eqnarray}\label{eq:b-analytic}
 b&=&\frac{1}{2}\,\left(1+\frac{\bar x}{\ell\sin\beta}\right)\,,
 \nonumber\\
 \sigma_b&=&\frac{L}{\ell}\,\frac{1}{2\sin\beta\sqrt{N}}\,,
\end{eqnarray}
where
\begin{equation}
 b=b_1=1-b_2\,.
\end{equation}
With $\sin\beta=0.259$ we have after one year $\sigma_b=0.053$. With
$b=0.5$ the $1\,\sigma$ uncertainty of each individual reactor
strength is $\pm10.7\%$, whereas the uncertainty of their sum is
only $\pm0.25\%$.

\subsection{Separation angle of reactors}

\label{sec:separation}

We have seen that for the Double Chooz setup, the directional
sensitivity of the near detector allows one to determine the
integrated source strength of the two reactors separately, even
though the uncertainty remains relatively large for one year of
data. We now ask the opposite question if one can separate the
``neutrino images'' of the two reactor cores, assuming their
relative strength is known, and assuming the detector
characteristics $L$ and $\ell$ have been established by other means
precisely enough that their uncertainty does not matter in the
following.

We primarily discuss how well the separation angle
$\phi=|\phi_1-\phi_2|$ can be determined, assuming we know that there
are exactly two sources in the $x$-$y$-plane that produce equal
numbers of events. The obvious observables are the central
coordinates $\bar x$, $\bar y$, and $\bar z$ of the displacement
vector distribution and their variances.

At first one may think that the width of the observed distribution
$f({\bf r})$ is broadened in the $x$-direction if one has two
sources because this distribution is a superposition of two Gaussian
distributions of width $L$ that are displaced relative to each other
by the distance $2\ell\sin\beta$ where we have assumed
$\phi_1=-\phi_2=\beta$. However, one easily finds in this case
\begin{equation}\label{eq:varxx}
 \langle x^2-\bar x^2\rangle=
 L^2+\ell^2\sin^2\beta\,.
\end{equation}
The rms width of the $x$-distribution increases only quadratically
in a small quantity with respect to the width $L$ of a single
source. For our parameters, the rms width of the double Gauss
function is the same as that of a single Gaussian within
$3\times10^{-4}$ and thus indistinguishable, even with five year's
data of almost a million events. Analogous conclusions pertain to
the higher moments of a single Gaussian compared to a double Gauss
function when their separation is much smaller than their width. In
other words, with the foreseen statistics of the Double Chooz near
detector, the neutrino images of the two reactors are far too
blurred to be separated.

However, it is still possible to distinguish a single source from
two sources if one takes advantage of the information encoded in the
average coordinates of the displacement vector distribution. In the
symmetric setup assumed here, information about the separation angle
is provided by the distribution $f_y$ that we have already used in
Sec.~\ref{sec:displacement} to determine $\ell$ if the separation
angle is known. Turning this argument around we may instead solve
for $\cos\beta$. Its uncertainty~is
\begin{equation}\label{eq:sigmacosbeta}
\sigma_{\cos\beta}=\frac{L}{\ell\,\sqrt{N}}\,,
\label{eq:sigcosb}
\end{equation}
which is $2.8\times10^{-2}$ for our usual parameters and one year's
data. If the separation angle is large, it can be ascertained with fairly
good accuracy. On the other hand, for the Double Chooz geometry with
$\cos\beta=0.966$, the reactors could be barely separated on this
basis, even with five year's data.  Moreover, since the angular
separation relies on a measurement of the quantity $\ell\cos\beta$, an
independent precise determination of $\ell$ is necessary.

In a certain number of cases the value for $\cos\beta$ implied by
the data will exceed unity and will thus be unphysical. In other
words, in these cases one cannot distinguish a single source from
two sources. Since $\cos\beta$ follows a Gaussian distribution and
using the width Eq.~(\ref{eq:sigmacosbeta}), this will be the case
with the ``confusion probability''
\begin{equation}\label{eq:confusion}
 p_{\rm confusion}=\frac{1}{2}\left[1+{\rm erf}\left(
 \frac{\cos\beta-1}{\sqrt{2}\,\sigma_{\cos\beta}}
 \right)\right]\,.
\end{equation}
For our usual parameters and one year's data we have $p_{\rm
confusion}=10.9\%$. After five years this number reduces to 0.29\%.
Turning this around, in more than 99.7\% of all cases the data will
imply the presence of two sources. Therefore, we estimate that the
Double Chooz near detector takes five years to distinguish the two
reactors from a single source with a $3\,\sigma$ confidence.

The reason for this relatively poor performance is that even for a
separation angle as large as $\phi=30^\circ$, the relevant quantity
$\cos(\phi/2)=0.966$ is difficult to distinguish from 1 for the
given statistics. On the other hand, if the separation angle is
somewhat larger, the deviation of $\cos(\phi/2)$ from 1 itself quickly
becomes of order unity so that $\cos(\phi/2)$ can be easily
distinguished from~1. Assuming the same detector characteristics and
$\phi=90^\circ$, we estimate that one could measure $\phi$ with a
precision of a few degrees even after only one year.

On the other hand, determining the central angle $\phi_c$ becomes
more difficult if the separation angle is too large.  For two
equally strong reactors on opposite sides of the detector
($\phi=180^\circ$), the separation angle can be very well measured,
whereas $\phi_c$ would remain completely undetermined. One would
conclude that there are two reactors at opposite sides without any
information on their absolute direction.

\section{Maximum-Likelihood Estimate}   \label{sec:maxlike}

\subsection{The method of maximum likelihood}

The analytic estimates of the previous section are based on
simple properties of the displacement-vectors distribution, notably
the coordinates of its center of gravity. Moreover, we have argued
that the shape of the double-Gauss distribution
Eq.~(\ref{eq:distribution}) is very similar to that of a single
Gaussian. Therefore, even for a five year exposure at the Double
Chooz near detector, the shape of the displacement-vector
distribution holds little additional information. Still, the maximum
information can be extracted by performing a maximum-likelihood
analysis, i.e., by fitting the measured distribution of displacement
vectors to a function of the form Eq.~(\ref{eq:distribution}).

The likelihood function of a set of $N$ independently measured
displacement vectors ${\bf r}_i=(x_i,y_i,z_i)$ is
\begin{equation}
L(\alpha)=\prod_{i=1}^{N}{f({\bf r}_i;\alpha)}\,,
\end{equation}
with $f({\bf r}_i;\alpha)$ given here by
Eq.~(\ref{eq:distribution}), where $\alpha$ denotes the a priori
unknown parameters $b_1$, $b_2$, $\phi_1$, and $\phi_2$. The set of
parameters that maximizes the likelihood returns the best-fit points
of a given data set. The maximum likelihood analysis is the most
powerful analysis method for unbinned data. Therefore, it is useful
to compare the analytic results of the previous section with the
maximum-likelihood method applied to sets of Monte Carlo data.

\begin{figure}[b]
\includegraphics[width=0.7\columnwidth]{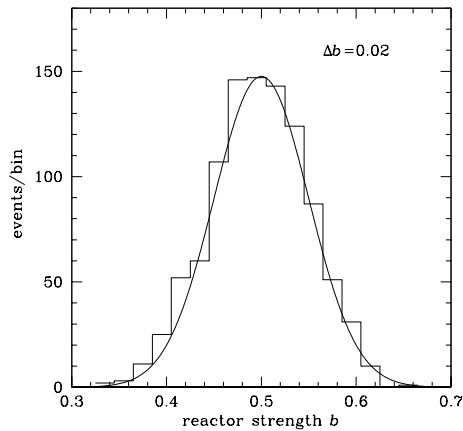}
\caption{Distribution of best-fit values of the reactor strength $b$
for 1,000 Monte Carlo realizations, assuming our usual parameters
and 1~year of data. The bin width is $\Delta b=0.02$. We also show a
Gaussian of width $\sigma_b=0.053$, representing the analytic
estimate of Eq.~(\ref{eq:b-analytic}).}\label{fig:b}
\end{figure}

\begin{figure}[b]
\includegraphics[width=0.7\columnwidth]{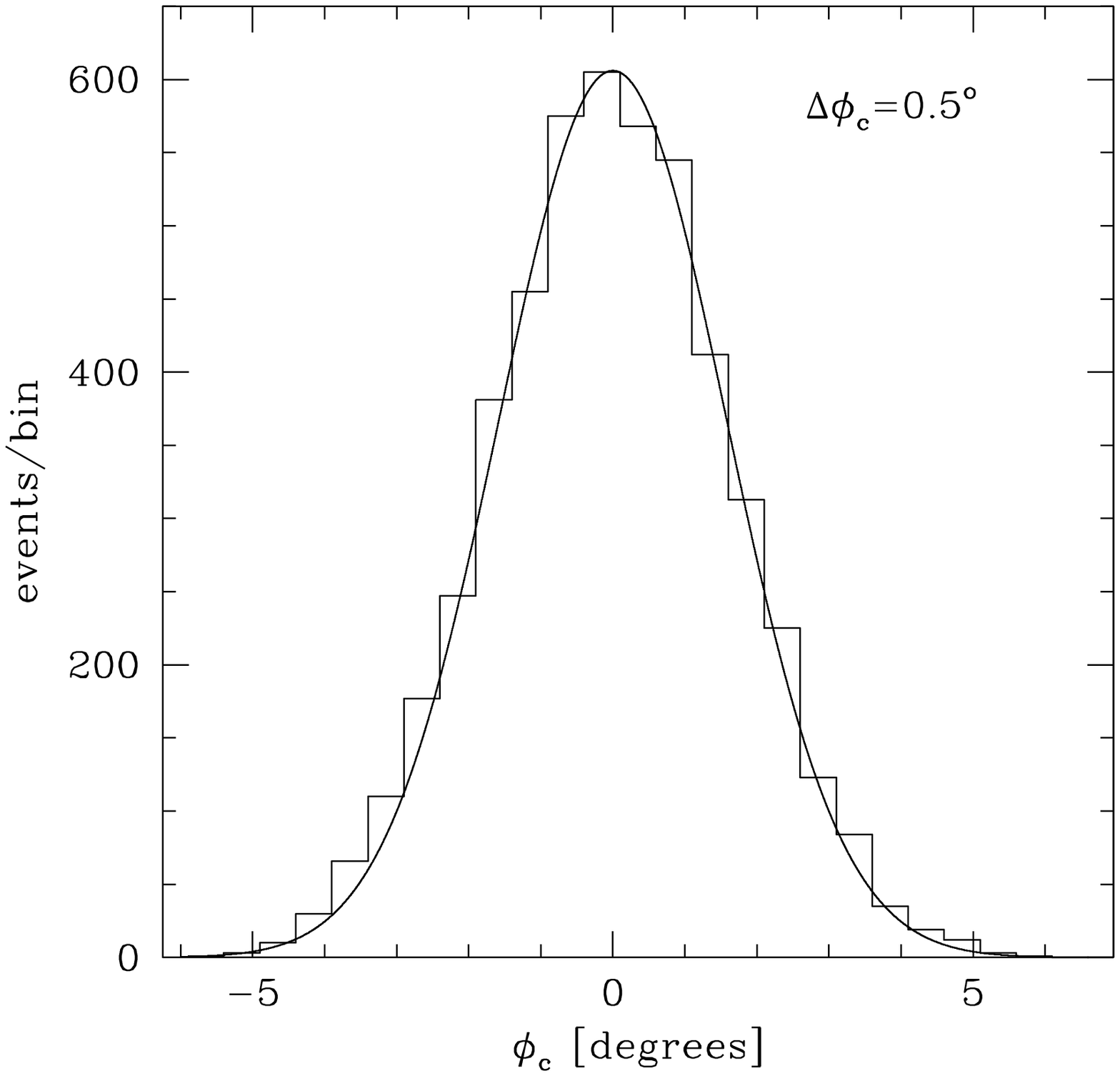}
\vskip12pt
\includegraphics[width=0.7\columnwidth]{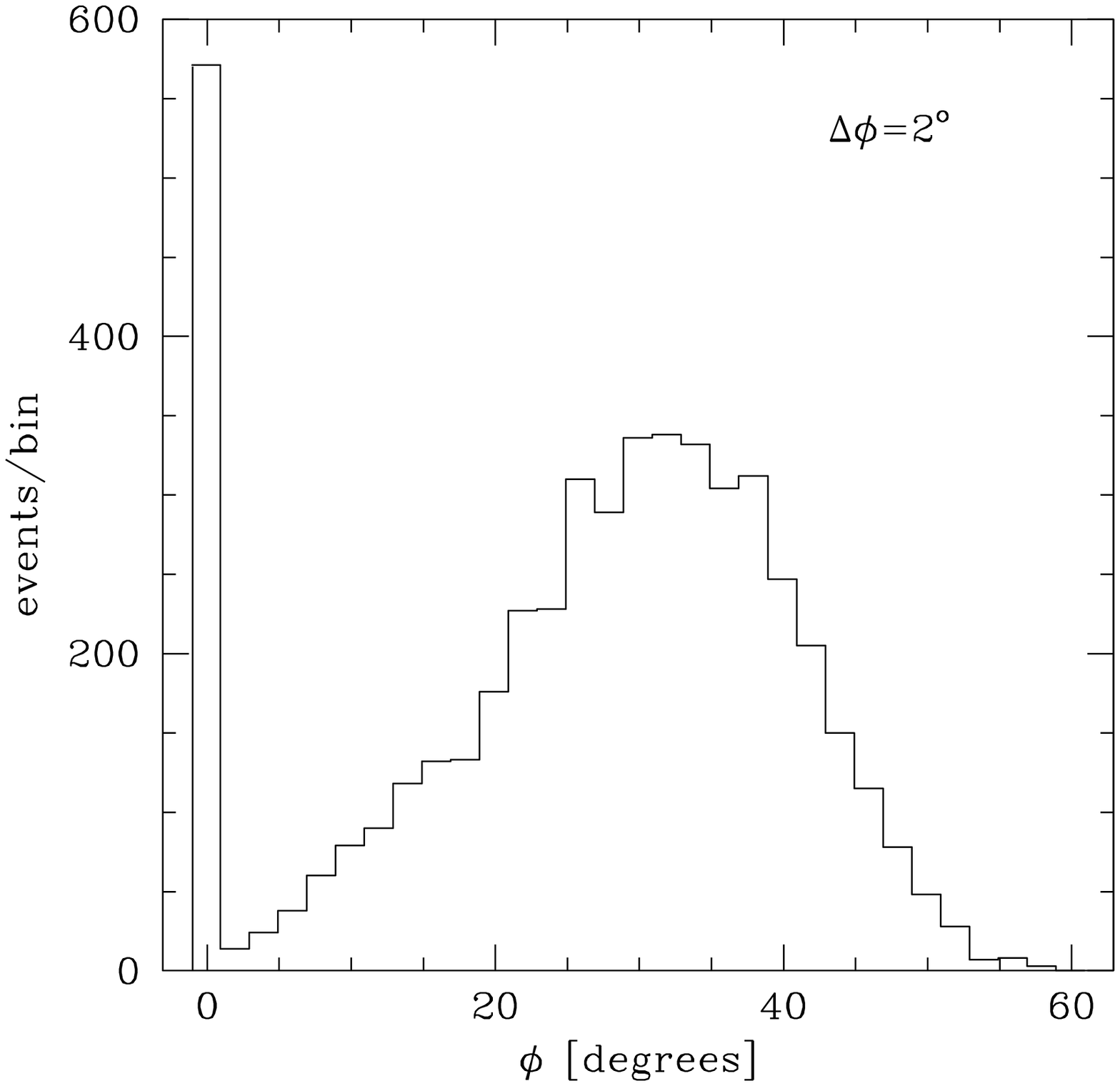}
\vskip12pt
\includegraphics[width=0.7\columnwidth]{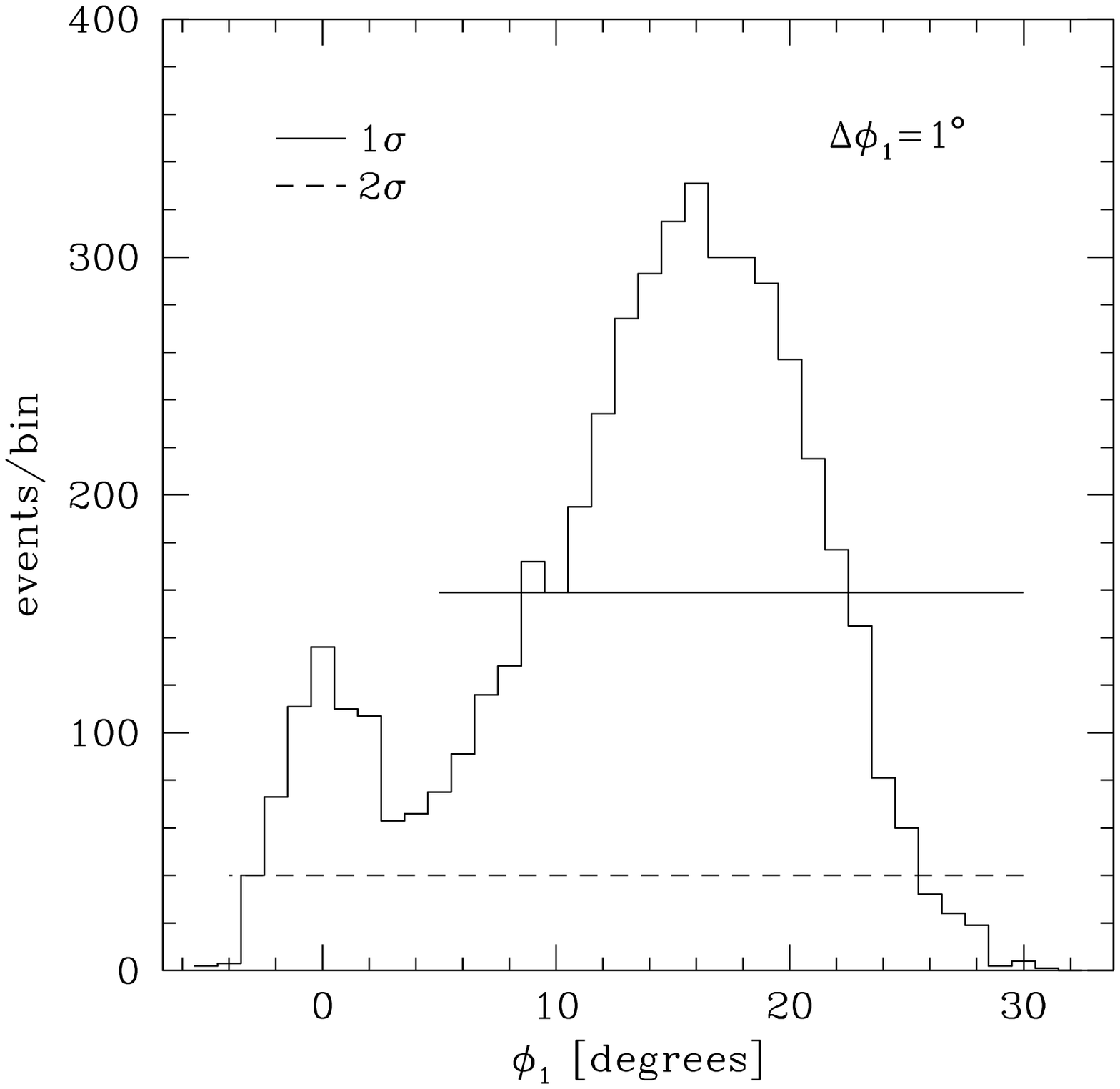}
\caption{Best-fit central angles $\phi_c=\phi_1+\phi_2$ ({\it top}),
separation angles $\phi=|\phi_1-\phi_2|$ ({\it middle}), and
azimuthal reactor location $\phi_1$ ({\it bottom}) for 5,000 Monte
Carlo realizations of our fiducial setup with $1.6\times 10^5$
events (one year). The different bin widths are indicated in each
panel. In the top panel we also show a Gaussian with the expected
width of $1.6^\circ$. In the bottom panel, the horizontal solid line
indicates the interval containing 68\% of all values, the dashed
line 95.4\%.}\label{fig:oneyear}
\end{figure}

\begin{figure}[b]
\includegraphics[width=0.7\columnwidth]{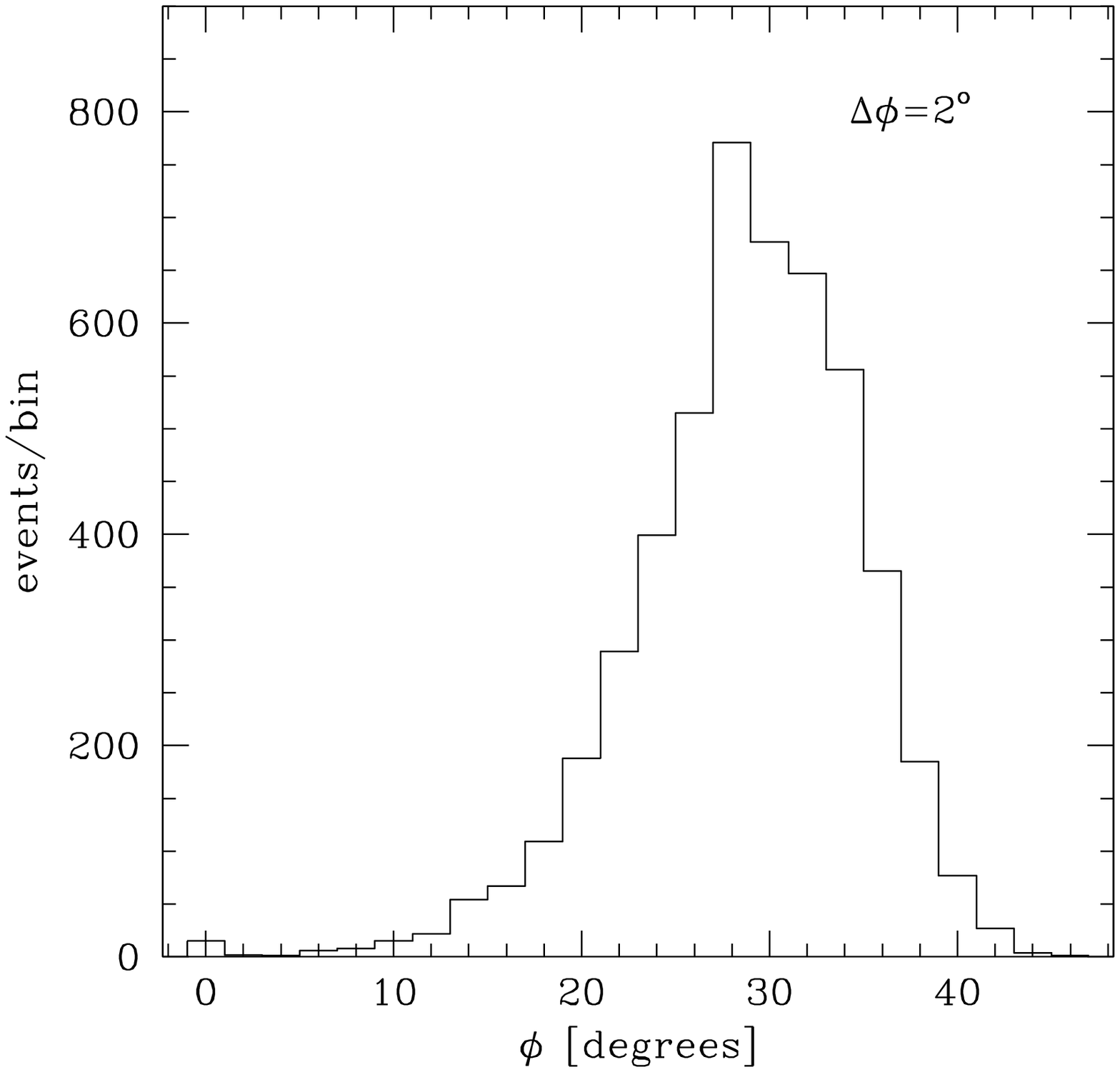}
\vskip12pt
\includegraphics[width=0.7\columnwidth]{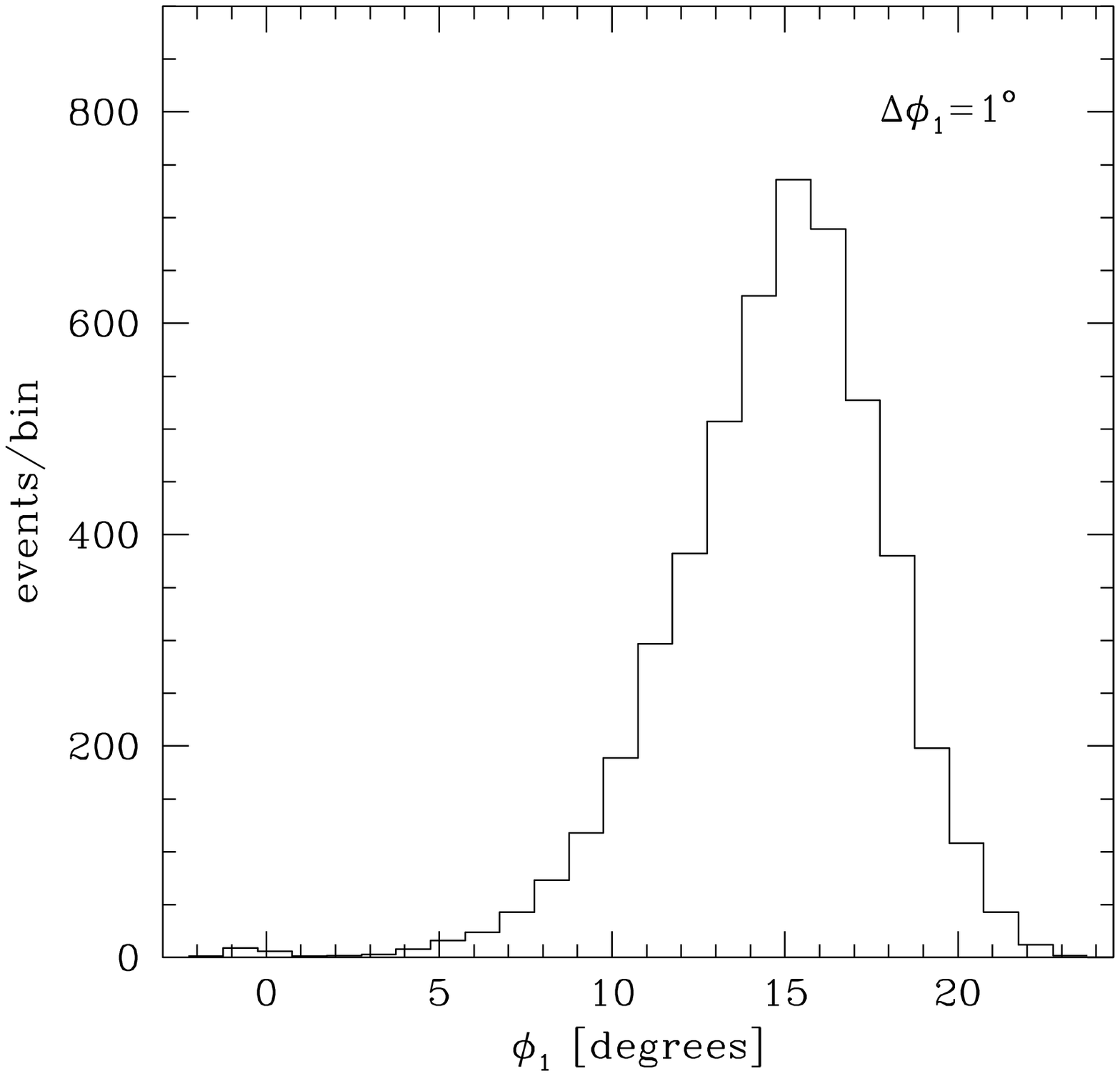}
\caption{Best-fit separation angles $\phi$ ({\it top}) and reactor
locations $\phi_1$ ({\it bottom}) as in Fig.~\ref{fig:oneyear}, here
for $N=8\times 10^6$ (five years).}\label{fig:sevenyear}
\end{figure}

\subsection{Relative reactor strength}

As a first example we return to the task of determining the relative
reactor strength. We assume that $\theta=0^\circ$ and
$\phi_1=-\phi_2=15^\circ$ is known. Under these circumstances the
distribution function factorizes and we will only keep $f_x(x)$
where we use $b_1=b_2=0.5$ to generate Monte Carlo data sets for $x$
with $N=1.6\times10^5$ events (one year). For each realization we
reconstruct $b$ by a maximum-likelihood fit. In Fig.~\ref{fig:b} we
show the distribution of best-fit values from 1,000 runs together
with a Gaussian distribution centered at $b=0.5$ and a width
$\sigma_{b}=0.053$ given by the analytic estimate
Eq.~(\ref{eq:b-analytic}). Both results correspond very well to each
other.

\begin{figure}[b]
\includegraphics[width=0.7\columnwidth]{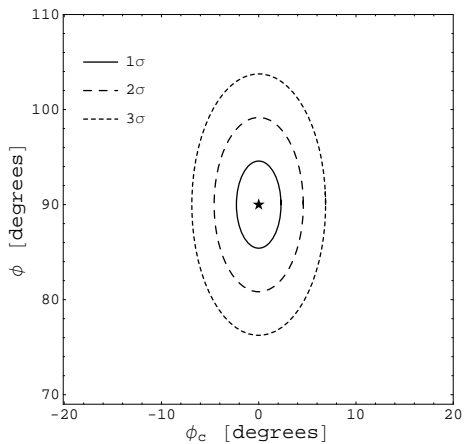}
\caption{Likelihood contours of $\phi$ and $\phi_c$ for one Monte
Carlo realization with $N=1.6\times 10^5$ events generated from a
setup with separation angle $\phi=90^\circ$ and central angle
$\phi_c=0^\circ$.} \label{fig:45}
\end{figure}

\subsection{Reactor directions}
\label{sec:maxl-dir}

As a next case we assume that the reactor strengths are known to be
$b_1=b_2=0.5$ and that the zenith angle for both sources is
$\theta=0^\circ$, whereas the azimuthal reactor locations $\phi_1$
and $\phi_2$ are our fit parameters. Since the $z$-distribution
factors out, we generate Monte Carlo data sets consisting of
$N=1.6\times10^5$ two-dimensional displacement vectors $(x,y)$. For
5,000 Monte Carlo realizations we show the distribution of
reconstructed best-fit central angles $\phi_c=b_1\phi_1+b_2\phi_2$
in the top panel of Fig.~\ref{fig:oneyear} together with a Gaussian
of width $1.6^\circ$ that corresponds to the expected analytic
width. Both results agree well with each other.

In the middle panel of Fig.~\ref{fig:oneyear} we show the
corresponding distribution of best-fit separation angles
$\phi=|\phi_1-\phi_2|$ which is taken to be a positive number
because the reactors have equal strength and thus are not
distinguishable. The distribution consists of a continuous component
and a spike at $\phi=0^\circ$. This solution corresponds to those
cases where the data prefer a single source as discussed in
Sec.~\ref{sec:separation}. According to Eq.~(\ref{eq:confusion})
this confusion should arise in 10.9\% of all cases, in good
agreement with the size of the spike in Fig.~\ref{fig:oneyear} if we
recall that the sum over all bins represents 5,000 Monte Carlo
realizations.

Finally we show in the bottom panel of Fig.~\ref{fig:oneyear} the
distribution of the reconstructed azimuthal reactor location
$\phi_1$. Since the two reactors are taken to have equal strength,
they are not distinguishable so that the distribution for $-\phi_2$
is the same. The distribution is bimodal with one peak at the true
location of $\phi_1=15^\circ$ and another at the central angle
$\phi_c=0^\circ$, corresponding to those cases where the two
reactors cannot be distinguished from a single source. The width of
this peak roughly corresponds to the width of the central-angle
distribution in the top panel. We have also indicated where 68\%
(solid line) and 95\% (dashed line) of all values fall around the
best-fit value. Even though the distributions are not Gaussian,
we refer to these regions as $1\,\sigma$, $2\,\sigma$ etc.\
intervals. The $1\,\sigma$ interval is approximately $12^\circ$
wide, whereas the $2\,\sigma$ interval includes the secondary peak
at~$0^\circ$. As the number of events increases, the $\phi_1$
distribution approaches a Gaussian and the peak at $0^\circ$
decreases. It takes roughly seven years of data to exclude this
secondary peak from the $3\,\sigma$ confidence region.

In Fig.~\ref{fig:sevenyear} we show the distribution of
separation angles $\phi$ and of reactor locations $\phi_1$ for 5,000
Monte Carlo realizations, each with 5 years of data ($N=8\times
10^5$). The spike at $0^\circ$ of the separation-angle distribution
has indeed decreased below 0.3\%, confirming our earlier analytic
estimate that with five year's data a $3\,\sigma$ separation of the
Double Chooz reactors is possible. Note, however, that the peak
around $0^\circ$ of the $\phi_1$ distribution remains in the
$3\,\sigma$ region.

As a more optimistic case, we consider a hypothetical setup with a
separation angle $\phi=90^\circ$. The distributions of the
reconstructed angles are essentially Gaussian in the relevant region
around the best-fit values. With $N=1.6\times10^5$, corresponding to
1~year at Double Chooz, the $1\,\sigma$ uncertainty for $\phi_1$ is
$\pm3.2^\circ$, for the central angle $\phi_c$ it is $\pm2.3^\circ$
and for the separation angle $\phi$ it is $\pm4.6^\circ$. We
illustrate this case in Fig.~\ref{fig:45} where we show the
likelihood contours for $\phi$ and $\phi_c$ corresponding to 1, 2,
and $3\,\sigma$ confidence regions. Note that the uncertainty of
$\phi$ is twice that of $\phi_c$ and that of $\phi_c$ is worse than
it was for a smaller separation angle.

\begin{figure}[b]
\begin{center}
\includegraphics[width=1.0\columnwidth]{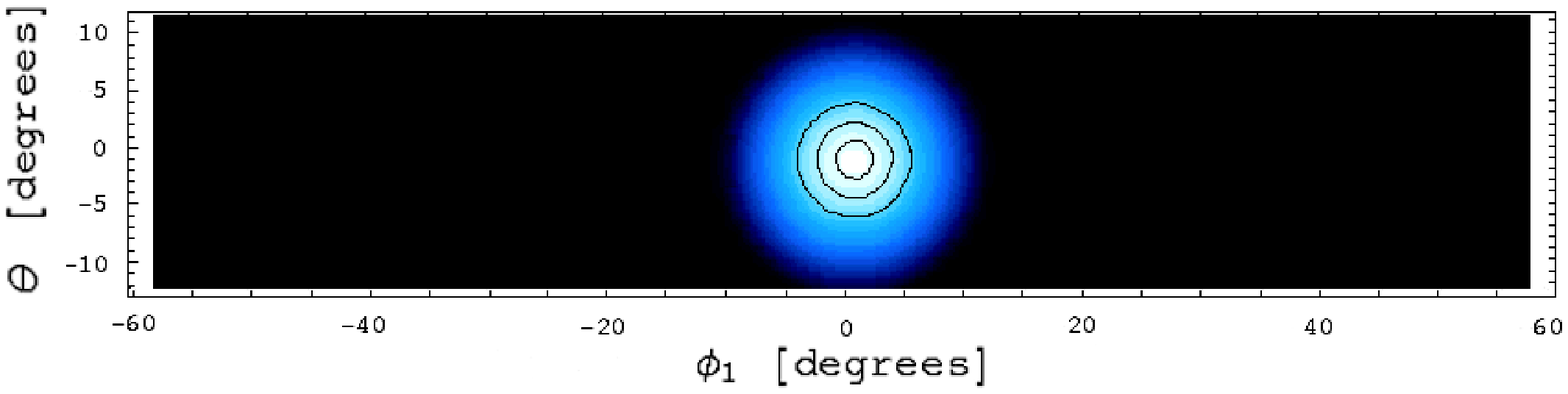}
\end{center}
\caption{Likelihood contours of $\phi_1$ and $\theta$ corresponding
to 1, 2 and $3\,\sigma$ confidence regions for a single reactor
source. This figure is based on one Monte Carlo realization with
$N=1.6\times10^5$ events.} \label{fig:uno}
\vskip12pt
\begin{center}
\includegraphics[width=1.0\columnwidth]{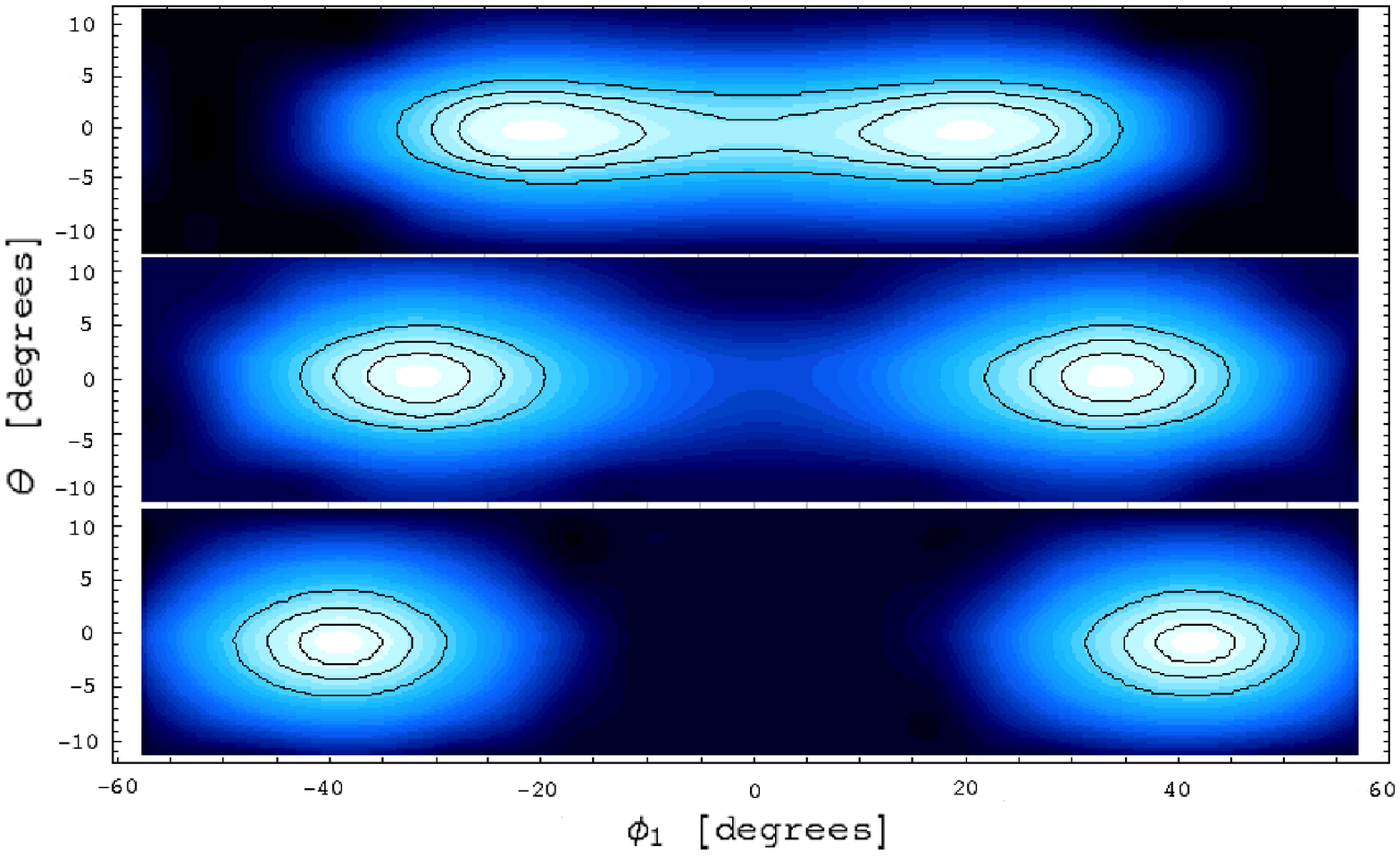}
\end{center}
\caption{Likelihood contours projected onto the $\phi_{1}$-$\theta$ plane, where the solid lines correspond to the 1, 2 and $3\,\sigma$ confidence regions. From top to bottom the true
separation angle was $30^\circ$, $70^\circ$, and $90^\circ$,
respectively. Each panel shows a ``typical'' Monte Carlo realization
consisting of $N=1.6\times10^5$ events.} \label{fig:trio}
\end{figure}

\subsection{Reactor directions with tilt}

\label{sec:tilt}

As a final example we include the zenith angle $\theta$ as a fit
parameter. In other words, we generate Monte Carlo data sets for the
full distribution function Eq.~(\ref{eq:distribution}) consisting of
$N=1.6\times10^5$ displacement vectors. As a first case we show in
Fig.~\ref{fig:uno} likelihood contours for $\phi_1$ and $\theta$
when there is a single source and the data are analyzed with the
prior assumption that indeed there is only a single source, i.e.,
assuming $phi_1=0^\circ$, $b_1=1$ and $b_2=0$. This figure can be taken as a
false-color neutrino image of a single reactor and illustrates the
single-source imaging power of the near detector at Double Chooz.
The solid lines correspond to the 1, 2 and $3\,\sigma$ contours.

Next we generate Monte Carlo realizations based on two sources
with separation angles $30^\circ$, $70^\circ$, and $90^\circ$,
respectively. The fit parameters of the maximum likelihood analysis
are $\theta$, $\phi_1$, and $\phi_2$, where both $\phi_1$ and
$\phi_2$ can a priori vary in the entire interval from $-180^\circ$
to $+180^\circ$. In Fig.~\ref{fig:trio} we show likelihood contours for three ``typical'' Monte Carlo realizations projected onto the $\phi_1$-$\theta$ plane. In our case of equal reactor strengths,
$L(\theta,\phi_1,\phi_2)=L(\theta,\phi_2,\phi_1)$ so that the
corresponding plot for $\phi_2$ is identical.

The panels of Fig.~\ref{fig:trio} can be taken as false-color
neutrino images of two reactors, although this interpretation must
be used with care because we show the likely location of one of the
reactors, not really the ``images'' of two neutrino sources.

For small separation angles, where the two ``reactor images'' merge,
the interpretation of the shown contours as 1, 2 and $3\,\sigma$
confidence regions is only approximate. We also note that the
distribution of zenith angles $\theta$, after marginalizing over the
azimuthal angles, is the same in all cases of Figs.~\ref{fig:uno}
and~\ref{fig:trio} within statistical fluctuations.

\section{Conclusions}                          \label{sec:conclusions}

We have investigated the ``neutrino imaging power'' of the Double
Chooz near detector that will collect as many as $1.6\times10^5$
events per year. Its angular sensitivity is based on the average
forward displacement of the neutron in the inverse-beta detection
reaction. For realistic assumptions derived from the properties of
the previous CHOOZ experiment, the width of the distribution of
reconstructed displacement vectors is about ten times larger than
the displacement itself, leading to an extremely blurred neutrino
image of a reactor.

For a single source, this image can be sharpened with enough
statistics so that its direction can be determined with very good
precision. At Double Chooz we obtain the average neutrino direction
with a $1\,\sigma$ half-cone aperture of $2.4^\circ$ with one year
of data.

However, with the given statistics, the images of two or more
sources merge completely in the sense that the shape of the
displacement-vector distribution is indistinguishable from that of a
single source. Yet the average of all measured displacement vectors
still contains nontrivial information on the source directions. It
is very difficult to separate two sources if their angular distance
$\phi$ is so small that $1-\cos(\phi/2)\ll1$. Thus for Double
Chooz we find that the setup is ineffectual for a clear separation
of the reactors. After 5 years of Double Chooz, the reactors can be
separated at the $3\,\sigma$ level.

With increasing separation angle it becomes much easier to
distinguish the reactors. For a setup with $\phi=90^\circ$, even one
year's data would be enough to measure the separation angle to about
$\pm10^\circ$ at $3\,\sigma$.

For Double Chooz, the location of the reactors is perfectly known,
of course. In this case one can use the angular sensitivity to
determine the two reactor strengths from the neutrino signal alone
and one can determine the detector response characteristics $L$ and
$\ell$ from the same data. At $1\,\sigma$, the one-year integrated
power of one of the reactors can be determined to $\pm11\%$, the
five-year integrated value to $\pm4.8\%$. The total event rate over
these periods is determined within $\pm0.25\%$ and $\pm0.11\%$,
respectively.


\acknowledgments
We thank J.~Kopp for useful discussions and L.~Oberauer for helpful
comments. This work was partly supported by the Transregio Sonderforschungsbereich TR27 ``Neutrinos and Beyond'' der Deutschen Forschungsgemeinschaft and by the European Union under the ILIAS project, contract No.~RII3-CT-2004-506222.


\end{document}